\newdimen\nude\newbox\chek
\def\slash#1{\setbox\chek=\hbox{$#1$}\nude=\wd\chek#1{\kern-\nude/}}

\def\to{\rightarrow}

\def\Sign{{\rm Sign}}

\def\mwhen{{\rm when}}
\magnification=1095
\parskip=-1pt plus 0.07 cm
\baselineskip= 14pt
 
\input psfig.sty
\rightline{KEK CP-057}
\vskip 1.0 cm
\centerline{{\bf Algorithm for the fermionic lines in 
GRACE-SUSY}\footnote{*}
{This work is partially supported by the Ministry of Education and 
Culture, Japan, under Grand-in-Aid for basic research program C
(no. 08640391).}}
\centerline{(revised version\footnote{**}
{ The original version is published in {\it Meiji Gakuin} 
Ronso, 594 (1997) 1.})}
\vskip 1.5 cm
\centerline{M. Kuroda}
\centerline{Institute of Physics, Meiji Gakuin University}
\centerline{Yokohama, Japan}
\centerline{and}
\centerline{Minami-Tateya Collaboration, KEK}
\centerline{Tsukuba, Japan}
\vskip 1.5 true cm

\noindent{\bf Abstract}\par
     Algorithm of constructing  Feynman amplitudes in the
framework of minimal supersymmetic extension of the standard model
is presented, which can be  easily implemented
in {\tt GRACE}, the  program of automatic generation of Feynman
amplitudes.  The equivalence of our method with the prescription given 
by Denner et al. is proved.
 
\vskip 1 truecm
      
\noindent{\bf 1.Introduction}\par
    The fundamental theory of elementary particle physics is 
described by lagrangians.  Given a lagrangian, the important task is to
calculate the reaction rates, such as production rates, scattering cross
sections or production cross sections of elementary particles, which
should be confronted with experimental data to confirm that the theory 
based
on the lagrangian is correct.  As the $S$ matrix is a nonlinear
functional of the lagrangian, we usually cannot calculate the reaction
rates exactly.  The conventional approach is, therefore, to use the
perturbative expansion of the $S$ matrix in the power of the coupling
constants and to calculate reaction rates to a certain order of coupling
constants.\par
     Given the order of perturbation, one draws all the Feynman diagrams
consistent with the theory and write down the corresponding Feynman
amplitudes.  The reaction rates are obtained by squaring the sum
of the amplitudes  and integrating over the phase space.  It is this
part of the calculation that is straightforward but
very tedious.  For example, for the process $e^+e^- \to W^+ W^-$, in the
lowest order in perturbation there are only three Feynman diagrams and
the computation of the cross section is rather easy.  Already at the
next order, however, there are as many as 140 Feynman diagrams (in
a covariant gauge) and the
calculation of the cross section by hand is a very hard task.\par
     Recently, several attempts have appeared[1][2][3][4] to 
automatize the
computation in the standard model of electroweak and QCD interactions.
 The most advanced one is the program called {\tt GRACE}, which
was developed by the Minami-Tateya Collaboration at KEK[5].  
The program is used, for example, to calculate
the cross section  of $e^+e^-\to f_1\bar f_2 f_3 \bar f_4$[6][7], which is
the background for the $W$-pair production process $e^+e^-\to W^+W^-$ and
consists of over 100 diagrams.\par
     In this paper I would like to give an instruction how to enlarge
and modify the {\tt GRACE} in order to implement the supersymmetric
theory (SUSY) in it.  The enlarged version of {\tt GRACE} is called
{\tt GRACE-SUSY}.  The motivation of SUSY is extensively discussed in the
literature (see, for example, ref.[8]) and the automatic calculation of
the SUSY processes is highly desired from several points of view.  First,
the theory of SUSY itself is very important, although at
present there is no evidence of the existence of SUSY particles in Nature.  
Secondly, for a given process with a given order of perturbation, there
are much more diagrams involved in SUSY than in the Standard Model (SM)
and the introduction of automatic computation of the amplitudes is
necessary when one wants to calculate various cross sections containing
SUSY particles.  Thirdly, expansion of the automatic computation of the
amplitudes from SM to SUSY is nontrivial due to the new aspects 
characteristic to SUSY and it is worthwhile to study how to embed
SUSY in {\tt GRACE} for its own sake.\par
     Since the new aspects of SUSY appear essentially in its fermionic
sector, the discussion is concentrated  
on the algorithm of how to treat fermionic lines in SUSY.
     In section 2, the algorithm for SM is briefly discussed, then in
section 3 its  extension to the SUSY case is derived.  Section 4 gives
some comments for the case of identical particles.  Some examples
are shown in section 5, which show that the algorithm and the rule of
{\tt GRACE-SUSY} are correct.  The equivalence of our method and
the prescription given by Denner et al. is proved in section 6.
Appendix A contains necessary
mathematical formulae of charge conjugation and Majorana particles.  
The rule and the algorithm for {\tt GRACE-SUSY} are
summarized in Appendix B and Appendix C.\par
\vskip 1 truecm

\noindent{\bf 2 Standard Model case}\par
     The main part of this section is developed by Minami-Tateya 
Collaboration, KEK[5].  
   We consider the process with $n_i$ incident spinor particles and $n_f$ 
outgoing spinor particles.\footnote{*}{  We use the terminology "spinor
particles" as the representative of fermions, antifermions and 
Majorana particles.}  The  matrix element is written as
$$
     <final\vert        {\cal L}(x_1).....{\cal L}(x_p) \vert initial>
     =<0\vert d_{n_i+n_f}....d_{ni+1} {\cal L}(x_1).....{\cal L}(x_p) 
                  d_{n_i}^\dagger....d_1^\dagger\vert 0>,
\eqno(1)
$$
where $d_i^\dagger$  can be either $a_i^\dagger$, the creation operator 
of fermion, or $b_i^\dagger$  the creation operator of antifermion.
Correspondingly, $d_i$  can be either $a_i$, the annihilation operator 
of fermion or  $b_i$, the annihilation operator of antifermion. 
At this stage 
it is very important to note that the order of creation operators 
and annihilation operators in (1) is chosen arbitrarily but once  
the order is fixed we have to stick to it, because any interchange of spinor
particles induces a minus sign. Tanaka and Kaneko [1][2] have chosen the
convention such that we assign
the particle numbers $1$ to $n_i$ to each of the 
$n_i$ incident spinor particles, namely Dirac 
fermions and  Dirac antifermions in this order. Likewise,
we assign the particle numbers $n_i+1$ to $n_i+n_f$ to each of the $n_f$ 
outgoing spinor particles, namely Dirac antifermions and  Dirac fermions  
in this order.  Note that
$$
     \vert n_i-n_f \vert = even,
\eqno(2)
$$
and therefore $n_i+n_f$ is also an even integer. \par
\medskip
 
\noindent{\bf 2.1 Pairing of spinors}\par
    Since the flow of spin one-half particles is not disconnected by the
interaction nor has any branch, any of the external spinor particles 
(independent of whether it belongs to the initial state or final state)
must be pairwise connected.  For a given process with $n_i+n_f$ spinor
particles, there exist ${{n_i+n_f}\over 2}$ spinor lines in each Feynman
diagram.
When one of such pairs  consists of two particles with particle number 
$m$ and $n$ with $m > n$, we write it symbolically as
$$
   (m,n),~~~~~~~~~ m > n.
\eqno(3)
$$
Using the notation of (3), pairing of $(n_i+n_f)$ spinor particles
is expressed as
$$
    P(a_1,a_2,.....,a_{ni+nf})\equiv
               (a_1,a_2)(a_3,a_4)....(a_{n_i+n_f-1},a_{n_i+n_f}).
\eqno(4)
$$
There are $(n_i+n_f-1)!!$ ways of combining 
$n_i+n_f$ particles into ${{n_i+n_f}\over 2}$ pairs. 
Of course, not all of the $(n_i+n_f-1)!!$ ways of pairing spinor particles
are physical or realized at the given order of the perturbation,
since fermion number must be conserved along the fermion lines.
For example, if both $m$ and $n$ belong to the initial
particles, one of them is fermion and the other must be
antifermion.
For each physically allowed $P$, there exists a distinct set of
Feynman diagrams.
\medskip
 
    Pairing of spinor particles means that the creation and annihilation 
operators appearing in (1) are rearranged such that the creation and/or
annihilation operators in a pair come next to each other.  Symbolically,
the original order of the operators becomes
$$
   (n_i+n_f, n_i+n_f-1, ......2,1) \to  \sum_P \Sign(P) P,
\eqno(5)
$$
where the sum is taken over $(n_i+n_f-1)!!$ possible ways of 
pairings $P$'s and $\Sign(P)$ is due to the permutation of 
creation and annihilation operators :
$$
       \Sign(P)= \cases{ +1, ~~~even,\cr
                        -1, ~~~odd, \cr }
\eqno(6)
$$
where {\it even} and {\it odd} mean that the series
$(a_1, a_2,......,a_{n_i+n_f})$ is obtained from the initial series 
$(n_i+n_f,,.......,2,1)$ (see (1)) by 
even permutation and odd permutation, respectively.  The $\Sign(P)$ is a
sign of the pairing $P$ relative to the trivial pairing,
$(n_i+n_f,n_i+n_f-1).....(4,3)(2,1)$.  Since only the relative phase 
is observable,
the absolute sign of the trivial pairing is irrelevant, and this rule
is applicable even when there is no physically allowed Feynman diagram 
for the trivial pairing.\footnote{*}{As one sees from (10)
a pair $(m,n)$ contains 
interaction lagrangians ${\cal L}(x_1).....{\cal L}(x_p)$.  Therefore in
permuting particles, some of the particles go through these lagrangians.
The interaction lagrangian being bilinear in spinor fields, there arises no
extra sign from shifting spinors through interaction lagrangians.}\par
\medskip
 
\noindent$\underline{\rm Example:~~ e^+e^- \to e^+e^- ~ in~ QED}$\par
$$
\matrix{    particle   &  e^-  & e^+  &  \to &  e^+   &  e^- \cr
            number     &  1    &  2   &      &  3     &  4   \cr}
\eqno(7)
$$
The possible $P$ and its sign is  
$$
\matrix{      P   &   (4,3)(2,1)  & (3,2)(4,1) & (4,2)(3,1) \cr
         \Sign(P)  &      +        &    +       &     -      \cr}
\eqno(8)
$$
and  the number of independent pairing is indeed $(2+2-1)!!=3$.
Among the three $P$'s, only the first two are allowed  and 
correspond, at tree level,  to the Feynman diagrams shown in fig.1.\par

\centerline{\psfig{file=fig1.epsf}}
\sevenrm 
\baselineskip = 8pt
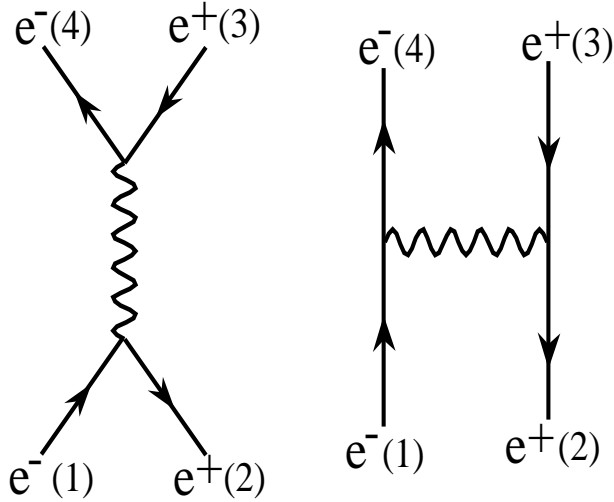
\noindent{Figure 1.  There are two Feynman diagrams for the process
$e^+e^- \to e^+e^-$ at tree level which correspond to the first two P's
in (8)}
\par
\tenrm
\baselineskip = 14pt

\medskip 

Recalling that the matrix element contains several interaction lagrangians 
(1) is now written as
$$
     <final\vert   {\cal L}(x_1).....{\cal L}(x_p) \vert initial>
   = \sum_P \Sign(P) \prod_{m>n}(m,n),
\eqno(9)
$$
where a  more exact meaning of the pair $(m,n)$ is given as follows:
$$
(m,n) = \cases{ <0\vert d_m {\cal L}...{\cal L} d_n^\dagger \vert 0>, 
              ~~~~~~~~~~\mwhen~m\in F,~~~n\in I,\cr
              <0\vert {\cal L}...{\cal L} d_m^\dagger d_n^\dagger \vert 0>, 
              ~~~~~~~~~~\mwhen~m,~n\in I,\cr
              <0\vert d_m d_n {\cal L}...{\cal L} \vert 0>, 
              ~~~~~~~~~~\mwhen~m,~n\in F,\cr}
\eqno(10)
$$
where
$$
     {\cal L} = \bar\Psi(x) \Gamma \Psi(x).
\eqno(11)
$$
Here $I$ and $F$ stand for the initial state and final state, respectively.
In (10), the first case corresponds to the scattering, the second case to the 
pair-annihilation and the last case to the pair-creation.  Note that due to
the restriction $m>n$, there exists no combination of the form 
$$
     <0\vert d_m^\dagger {\cal L}...{\cal L} d_n \vert 0>.
\eqno(12)
$$
\vskip 1 truecm

\noindent{\bf 2.2  Grace lines}\par
   In this subsection, we introduce the concept of 
{\it Grace line}, {\it Grace line amplitude}, 
{\it M-direction}  and {\it F-direction}  which are
useful in constructing  Feynman amplitudes from the pairs $(m,n)$.
As (10) shows, a pair $(m,n)$ with $m>n$ corresponds to 
a spinor line in a certain 
Feynman diagram.  Let us stretch such a spinor line in a straight
line and display it horizontally.  Such a horizontal
line is called {\it Grace spinor line}, in short {\bf Grace line}, 
which is denoted by $G_{p,q}$ when
he particle numbers of the external spinor particles at the left end 
and at the right end are $p$ and  $q$, respectively.  For a given
pair $(m,n)$, there exist two {\it Grace lines}, $G_{m,n}$ and $G_{n,m}$,
depending on which particle is put on the right and which on the left.\par
    In order to restore the one-to-one correspondence between $(m,n)$ and
the {\it Grace line}, we {\it assign}  $G_{m,n}$ to $(m,n)$ and $G_{n,m}$ 
to $-(m,n)$.  In fact, for annihilation and pair-creation, one can 
interchange two creation or annihilation operators in (10), leading 
to $-(m,n)$.
Therefore, it is natural to extend $(m,n)$ to the case when $n>m$ by
the following equation:
$$
         (m,n)=-(n,m), ~~~~~\mwhen ~~n>m.
\eqno(13)
$$
For annihilation and pair-creation, (13) is consistent with the definition 
of $(m,n)$, (10),
while for scattering, $(n,m)$ with $m>n$ is defined 
only through (13).  When $d_m$ and $d_n^\dagger$ are interchanged naively
in (10), $(n,m)$ vanishes, contradicting with (13).  \par
    Our task is to find a rule of writing down,  for a given {\it Grace line} 
$G_{m,n}$, the corresponding expression. We call such an 
expression {\it Grace spinor line amplitude}, in short {\bf Grace line 
amplitude}, and denote it by $[G_{m,n}]$.  The {\it Grace line 
amplitude} is constructed by replacing each block of the Grace line, the 
external spinor wave functions, vertices and propagators,  by its
representation and putting it exactly at the place where it appears in 
the Grace line.  The {\it Grace line amplitude} is
the faithful representation of the {\it Grace line}.  The {\it Grace line 
amplitude}, $[G_{m,n}]$,  starts with  the external spinor wave function 
corresponding to the particle $n$ at the right endpoint and ends with the 
external spinor wave function corresponding to the particle $m$ at the 
left endpoint.\par
     Using the generalized 
definition of $(m,n)$, the correspondence among the pair $(m,n)$, the 
Grace line $G_{m,n}$ and Grace line amplitude $[G_{m,n}]$ can 
be expressed as 
$$
        (m,n)  \leftrightarrow  G_{m,n} \leftrightarrow [G_{m,n}].
\eqno(14)
$$
Therefore, the rule  of Grace line 
amplitude must respect the property (14) and  satisfy
an important requirement,  
$$
         [G_{m,n}]=- [G_{n,m}].
\eqno(15)
$$
\medskip
      A {\it Grace line} has $k$ vertices and $k-1$ spinor propagators 
when there are $k$ interaction lagrangians in $(m,n)$ (see (10)).
On each segment of the spinor line separated by the vertices,
we assign two directions,
the momentum direction ({\bf M-direction}) and the fermionic 
direction ({\bf F-direction}).  Tanaka[2] named the direction from right
to left $(\leftarrow)$  as the {\bf A-direction}
(amplitude direction), 
which is the direction in which the spinors
are ordered in the Feynman amplitude when a fermion comes in and goes out 
after some interactions in between.\par
\medskip
\vfill\eject 
 
\noindent{\bf momentum direction}\par
     The momentum direction (M-direction) is assigned in a natural way.  
For the external incoming particles the M-direction 
is the direction towards the vertex, while for the external outgoing 
particles the M-direction is the direction from the vertex.
For the internal fermions, the M-direction is always taken in  
the A-direction (irrespective of whether the internal propagator is 
that of fermion or antifermion when seen in the 
A-direction).\par
\medskip
 
\noindent{\bf fermionic direction}\par
     For the external Dirac particle,
we can assign the F-direction in a natural way:  for fermions the 
F-direction is in the M-direction and for antifermions in 
the opposite direction to its M-direction.
Since the fermion number is conserved in the Standard Model, external
particles at the both ends of the Grace line must have the same
F-direction.  This fact uniquely fixes the F-direction of the internal
fermion propagators:  The F-direction of the internal propagators must
be in the same direction as that of the external particles. Therefore,
in the Standard Model we can define the F-direction for each Grace line
itself.\par
     For illustration purpose, we show $(m,n)$ and the corresponding
general form of Grace lines, $G_{m,n}$ in Table 1, where $"\bullet"$
stands for the interaction vertex. The fermion propagators between
two such vertices are not explicitly shown.,\par

\topinsert
{\offinterlineskip \tabskip=0pt
\halign{ \strut
    \vrule# &   \quad # \quad &
    \vrule# &   \quad # \quad &
    \vrule# &   \quad # \quad &
    \vrule# \cr
\noalign{\hrule}
  & pair~annihilation && $(b_m^\dagger a_n^\dagger)$  && 
   $\longleftarrow\bullet\bullet\bullet\bullet\bullet\longleftarrow$ & \cr
  &  &&  && $p_m\rightarrow~~~~~~~~~~\leftarrow p_n$     & \cr
\noalign{\hrule}
  & pair~creation     && $(a_m b_n)$   & &
   $\longleftarrow\bullet\bullet\bullet\bullet\bullet\longleftarrow$  & \cr
  &  &&  && $p_m \leftarrow~~~~~~~~~~\rightarrow p_n$ &\cr
\noalign{\hrule}
  &  scattering        && $(a_m a_n^\dagger)$          && 
   $\longleftarrow \bullet\bullet\bullet\bullet\bullet\longleftarrow$ & \cr
  &  &&   && $p_m\leftarrow~~~~~~~~~~\leftarrow p_n$ & \cr
\noalign{\hrule}
  &  scattering        && $(b_m b_n^\dagger)$          && 
   $\longrightarrow \bullet\bullet\bullet\bullet\bullet\longrightarrow$ & \cr
  &  &&   && $p_m\leftarrow~~~~~~~~~~\leftarrow p_n$ & \cr
\noalign{\hrule}  }}
\medskip
\noindent{Table 1:}  Assignment of F-direction(shown on the Grace 
line by arrow) and 
M-direction(shown in the lower part) to the external spinor particles.
Here $a, a^\dagger$ refer to fermions and $b, b^\dagger$ refer 
to antifermions.  $"\bullet"$ 
stands for the interaction vertex and the fermion propagators between
two such vertices are not explicitly shown.\par
\vskip 1 truecm
\endinsert

\noindent{\bf 2.3. Rule of $[G_{m,n}]$}\par           
     In this subsection, I will find the {\it Grace line rule}, the  rule of 
writing a Grace line amplitude $[G_{m,n}]$ from the corresponding 
{\it Grace line} $G_{m,n}$.  It should be stressed that the 
{\it Grace line rule} is not identical to the Feynman rule.  \par
     The direct calculation of (11) gives
$$
\eqalignno{
   <0\vert{\cal L}(x)...{\cal L}(z)b_m^\dagger a_n^\dagger\vert 0> =&
                      -\bar v_m \Gamma(x) S...\Gamma(z) u_n,    &(16a)\cr
   <0\vert a_m b_n{\cal L}(x)...{\cal L}(z)\vert 0> =&
                      -\bar u_m \Gamma(x) S...\Gamma(z) v_n,    &(16b)\cr
   <0\vert a_m{\cal L}(x)...{\cal L}(z) a_n^\dagger\vert 0> =&
                      +\bar u_m \Gamma(x) S...\Gamma(z) u_n,    &(16c)\cr
   <0\vert b_m{\cal L}(x)...{\cal L}(z)b_n^\dagger\vert 0> =&
                      -\bar v_n \Gamma(x) S...\Gamma(z) v_m \cr
            =&-v_m^T\Gamma(z)^T ....S^T\Gamma(x)^T (\bar v_n)^T,&(16d)\cr
}$$
where $\Gamma$ and $S$ stand for the vertex and the spinor propagator,
respectively.  The Grace lines given in Table 1 correspond 
exactly to each of the expressions (16).  Note that all
these expressions (16) correspond to a single notation 
$G_{m,n}$ and the Grace line rule must automatically
reproduce all of four equations given in (16), depending on the nature 
of $m$ and $n$ (creation operator
or annihilation operator; or fermion or antifermion).\par
     Comparing (16) and Table 1, one immediately obtains the first rule 
of the Grace line amplitude  concerning the external spinors at the 
ends of the Grace lines,
which is summarized in Table 2.  In order to explain the minus signs
in (16a) and (16b) relative to (16c), one has to supplement the
Table 2  with the following rule:\par
\medskip
\noindent{$\underline{\bf Rule~1:}$}  Multiply the Grace spinor amplitude 
by ($-$) for each
external spinor whose M-direction is opposite to the A-direction.\par
\medskip
\noindent{This} minus sign could, in principle, be put in the definition 
of endpoint spinors given in Table 2. \par

\topinsert
{\offinterlineskip \tabskip=0pt
\halign{ \strut
   \vrule#& \quad # \quad &
   \vrule#& \quad # \quad &
   \vrule#& \quad # \quad &
   \vrule#& \quad # \quad &
   \vrule#& \quad # \quad &
   \vrule#  \cr
\noalign{\hrule}
  & F-direction  && $\longleftarrow\bullet$ &&  $\longrightarrow\bullet$ &&
                    $\bullet\longleftarrow$ &&  $\bullet\longrightarrow$ & \cr
  &  M-direction &&                 &&               && &&              & \cr
\noalign{\hrule}
  & $\longleftarrow$  &&   $\bar u(p)$   &&    $ v(p)^T$  
                      &&   $u(p)$        &&    $ \bar v(p)^T$ & \cr
  & $\longrightarrow$ &&   $\bar v(p)$   &&    $ u(p)^T$  
                      &&   $v(p)$        &&    $ \bar u(p)^T$ & \cr
\noalign{\hrule} }} 
\medskip
\noindent{Table 2:} Spinor assignment at the  left 
endpoint and at right endpoint.
\endinsert

   Next, we turn to the vertices and propagators.
     We adopt the convention that 
when the F-direction of Grace line is  
the same  as the A-direction, we use the usual spinor propagator $S(k)$  and
the usual vertex  defined by the model lagrangian.\footnote{*}{Here we tacitly
assume that the model lagrangian which defines $\Gamma_{AB}$ etc 
is written in terms of particles(e.g.
electron) but not in terms of antiparticles (e.g. positron).  This seems
to be a trivial statement, but actually it is nontrivial as we soon
notice when we apply the Grace line rule to SUSY in which whether a 
certain particle is a fermion or antifermion is a matter of convention.}
More explicitly stated, for the model lagrangian
$$
     {\cal L}= \bar\Psi_A \Gamma_{A,B} \Psi_B +  
               \bar\Psi_B \Gamma_{B,A} \Psi_A,
\eqno(17)
$$
the propagator and the vertex function of the Dirac spinor in the 
Grace line amplitude are given by
$$
\eqalign{
        \bullet \longleftarrow \longleftarrow\bullet~~~~~~~~~~~~  & =
      ~~~~~~~~  S(k)\cr
    ~~~~~ \leftarrow k ~~~~~~~~~~~~~~~~& \cr
     A~ \longleftarrow\bullet\longleftarrow~B~~~~~~~~ &=
      ~~~~~~~~ \Gamma_{A,B} \cr
     B~ \longleftarrow\bullet\longleftarrow~A~~~~~~~~ &=
      ~~~~~~~~ \Gamma_{B,A} \cr
}\eqno(18)
$$
Here $k$ denotes
the propagator momentum, which by definition of our assignment of
M-direction is along the A-direction, and in turn in the present case,
in the F-direction of the Grace line.  Then, one can easily write 
down the Grace line amplitude $[G_{m,n}]$ corresponding to the first 
three Grace lines depicted in Table 1,
with the result which is indeed identical to the right hand side of 
(16a),(16b) and (16c).  
When the F-direction is opposite to the 
A-direction as is the case in (16d)  we use the following rule.
\medskip
\noindent{$\underline{\bf Rule~2:}$}  When the F-direction of the 
Grace line is opposite 
to the A-direction, the propagator and the vertex function of the 
Grace spinor amplitudes are given by $-S^T(-k)$ and $-\Gamma^T$, 
respectively, where $k$ is the momentum of the propagator along the
A-direction (Note that the momentum of the internal spinors is always
defined along the A-direction).  Schematically,
$$
\eqalign{
        \bullet \longrightarrow \longrightarrow\bullet~~~~~~~~~~~~  &
       = ~~~~~~~~-S^T(-k)\cr
    ~~~~~~~ \leftarrow k ~~~~~~~~~~~~~~~~~~& \cr
     B~ \longrightarrow\bullet\longrightarrow~A ~~~~~~~~&
       = ~~~~~~~~-\Gamma^T_{A,B} \cr
     A~ \longrightarrow\bullet\longrightarrow~B ~~~~~~~~&
       = ~~~~~~~~ -\Gamma^T_{B,A} \cr
}\eqno(19)
$$
\medskip
     Table 2 and Rules 1, 2 are all we need for writing down
the Grace line amplitude for a given Grace line in the Standard Model.  
\medskip

To be complete, we will check that the above rules indeed respect
the antisymmetry property (15).  When the direction of the Grace lines 
depicted in Table 1 is reversed together with their momenta, 
we obtain the Grace lines, $G_{n,m}$.  
Applying the rule of Grace line amplitudes to $G_{n,m}$
the Grace line amplitudes of the reversed Grace lines
are obtained with the following result:
$$
\eqalignno{
   \longrightarrow\bullet\bullet\bullet\bullet\bullet\longrightarrow~~~~~~~~~~  
           & = -u_n^T(-\Gamma^T)(-S^T)...(-\Gamma^T)(\bar v_m)^T  & \cr
               p_n\rightarrow~~~~~~~\leftarrow p_m ~~~~~~~~
           & = \bar v_m \Gamma S ...\Gamma u_n,                   &(20a)\cr
  \longrightarrow\bullet\bullet\bullet\bullet\bullet\longrightarrow~~~~~~~~~~   
           & = -v_n^T(-\Gamma^T)(-S^T)...(-\Gamma^T)(\bar u_m)^T  & \cr
               p_n\leftarrow~~~~~~~\rightarrow p_m ~~~~~~~~
           & = \bar u_m \Gamma S ...\Gamma v_n,                   &(20b)\cr
  \longrightarrow\bullet\bullet\bullet\bullet\bullet\longrightarrow~~~~~~~~~~  
           & = (-)(-)u_n^T(-\Gamma^T)(-S^T)...(-\Gamma^T)(\bar u_m)^T &  \cr
               p_n\rightarrow~~~~~~~\rightarrow p_m ~~~~~~~~
           & = -\bar u_m \Gamma S ...\Gamma u_n,                  &(20c)\cr
   \longleftarrow\bullet\bullet\bullet\bullet\bullet\longleftarrow~~~~~~~~~~   
           & = (-)(-)\bar v_n\Gamma S...\Gamma v_m                & \cr
               p_n\rightarrow~~~~~~~\rightarrow p_m ~~~~~~~~
           & = \bar v_n \Gamma S ...\Gamma v_m,                   &(20d)\cr
}$$
Comparing (20) with (16), one sees that  the requirement 
$[G_{m,n}]=-[G_{n,m}]$ is indeed satisfied.\par
\medskip

     The Feynman amplitude corresponding to the Feynman diagram
characterized by the partition $P$ is given by the product of
${{n_i+n_f}\over 2}$ Grace line amplitudes $[G_{m,n}]$ and the
appropriate bosonic parts(purely bosonic vertices and bosonic
propagators) which bridge vertices on the Grace lines.

\vskip 1 truecm

\noindent{\bf 3 SUSY case}\par
   We restrict our discussion to the minimal SUSY extension of the standard
model (MSSM).  It contains several new particles which do not exist in
SM.  They are\par
\leftskip +2cm
   charged Higgs boson, $(H^\pm)$,\par
   three neutral Higgs bosons (two $CP$ even states and 
   one $CP$ odd state),\par
   two charginos, $\tilde\chi_1^\pm$, $\tilde\chi_2^\pm$,\par
   four Majorana neutralinos, $\tilde \chi_1^0$, $\tilde \chi_2^0$,
                              $\tilde \chi_3^0$, $\tilde \chi_4^0$, \par
   7 right handed scalars and 8 left handed scalars. \par
\leftskip 0 cm
\medskip
For the number of the scalars  cited above the colour factor is taken 
into account.\par
       What is new in SUSY, in addition to the new particles, is
the existence of the fermion number violating interactions which come
about from two kinds of interactions.  Neutralino is a Majorana
particle which behaves itself as fermion as well as as antifemion.
Therefore, neutralino propagator does not necessarily conserve fermion
number.  Secondly, chargino behaves as fermion at some interaction
vertices but as antifermion at some other vertices.  Therefore,
chargino can flip the fermion number at certain vertices.  Note that,
consequently, the fermion number of charginos is not determined by their
interaction.  We {\it define} in our convention that the {\it positive}
charginos are Dirac fermions.\par
     In the MSSM case, therefore, the F-direction is not necessarily 
conserved along the GRACE line and the clash of the F-directions happens
either at vertices involving Majorana particle(s) or chargino(s), 
or at Majorana spinor propagators.  Before going to the next step, we
have to recall several important properties of Majorana particles which
are useful for the ensuing discussion.  More details are discussed in 
Appendix A.\par
  
\vskip 1 truecm
 
\noindent{\bf 3.1 Majorana particle}\par
     Majorana particle behaves as particle as well as antiparticle:
$ \Psi = \Psi^c\equiv C\bar\Psi^T$.
As our convention, we fix the relative 
phase of the spinor wave functions such that 
$$
       C{\bar v(k,\lambda)}^T = u(k,\lambda),~~~~~~~~~~
       C{\bar u(k,\lambda)}^T = v(k,\lambda),
\eqno(21)
$$
where $C$ stands for the charge conjugation matrix, which satisfies,
independent of the representation of gamma matrices, the following conditions,
$$
   C^\dagger = C^{-1},~~~~~
   C =  -C^T,~~~~~~~~C\gamma_\mu^T C^{-1} = -\gamma_\mu.
\eqno(22)
$$
     For Majorana particles,  there are three kinds of propagators,
$$
\eqalignno{
    <\Psi \bar\Psi> =& ~~~\bullet\longleftarrow\longleftarrow\bullet ~~~
                    = {i\over{\slash k - m}},   &(23a) \cr    
    <\bar\Psi \bar\Psi> =& ~~~\bullet\longrightarrow\longleftarrow\bullet ~~~
                    = C^{-1}{i\over{\slash k - m}},   &(23b) \cr    
        <\Psi \Psi> =& ~~~\bullet\longleftarrow\longrightarrow\bullet ~~~
                    = {i\over{\slash k - m}}C^T.   &(23c) \cr   
}$$
In (23) the momentum $k$ runs from right to left. 
The interaction lagrangian at Majorana-Dirac vertices is given as
$$
   {\cal L}_{MD} = \bar\Psi(M_\alpha)\Gamma_{M_\alpha,D_\beta} \Psi(D_\beta)
           +  \bar\Psi(D_\alpha)\Gamma_{D_\alpha,M_\beta} \Psi(M_\beta) 
\eqno(24)
$$
with $\Gamma_{D_\alpha,M_\beta}=\gamma_0\Gamma_{M_\beta,D_\alpha}^\dagger\gamma_0$.
The Majorana-Majorana interaction is given by 
$$
   {\cal L}_{MM} = \sum_{\alpha,\beta}
                   \bar\Psi_L(M_\alpha) \tilde \Gamma_{\alpha,\beta}
                   \Psi_L(M_\beta).
\eqno(25)
$$
Using the Majorana condition, which states
$$
\eqalign{
  C(\bar\Psi_{\alpha L})^T =& (\Psi_{\alpha L})^c = \Psi_{\alpha R}, \cr
 (C^{-1}\Psi_{\beta L})^T =& (\bar\Psi_{\beta L})^c = \bar\Psi_{\beta R},
}\eqno(26)
$$
the lagrangian (25) can be recast in the following form, which can be
directly applicable to the Feynman rule,
$$
   {\cal L}_{MM} = \sum_{\alpha\ge\beta}
                   \bar\Psi(M_\alpha)\Gamma_{\alpha,\beta}\Psi(M_\beta).
\eqno(27)
$$\par
 
\vskip 1 truecm
 
\noindent{\bf 3.2 Grace line rule for MSSM}\par
     We will modify the Grace line rule of section 2 so that
Majorana particles as well as the F-direction clashing vertices are
included in the rule.\par
     We assign the particle number $1$ to $n_i$ to each of the $n_i$
incoming spinor particles, namely Dirac fermions, Dirac antifermions and
Majorana particles in this order.  Likewise, we assign the particle
number $n_i+1$ to $n_i+n_f$ to each of the $n_f$ outgoing spinors, namely
Dirac antifermions, Dirac fermions and Majorana particles in this
order.\par
     The M-direction of the Majorana particle is assigned, like the
conventional fermions, to its momentum direction for external 
Majorana particles and to the A-direction for internal Majorana
particles.  For Majorana particles which
are self charge-conjugate, it is irrelevant whether one assigns the
F-direction along the A-direction or opposite to the A-direction.  For
our convention, we define the F-direction of Majorana particles by
their M-direction, which is along the A-direction for  internal
Majorana particles.  This corresponds to
regarding all the Majorana particles as fermion.  Therefore, only the
first type of the propagator shown in (23) appears in our Grace line rule
and the clashing of the F-direction occurs only at vertices.\par
     For the lagrangian (24),  we assign the vertex function as follows
(The irrelevant constant $i$ is not explicitly exhibited hereafter),
$$
\eqalign{
    M_\alpha \longleftarrow \bullet \longleftarrow D_\beta &~~~~~~~~~~
                 \Gamma_{M_\alpha, D_\beta},\cr
    M_\alpha \longrightarrow \bullet \longleftarrow D_\beta &~~~~~~~~~~
                 (C^{-1})^T\Gamma_{M_\alpha, D_\beta},\cr
    D_\alpha \longleftarrow \bullet \longleftarrow M_\beta &~~~~~~~~~~
                 \Gamma_{D_\alpha, M_\beta},\cr
    D_\alpha \longleftarrow \bullet \longrightarrow M_\beta &~~~~~~~~~~
                 \Gamma_{D_\alpha, M_\beta}C.\cr
}\eqno(28)
$$
For the vertex which has an opposite F-directions, we apply the rule 2
and obtain the following rule.\footnote{*}{This can be seen by considering
the processes whose external particles are conventional Dirac fermions 
and in which Majorana particles appear only in the internal lines.   In such
processes, the argument for the SM given in the section 2 is applicable.}
$$
\eqalign{
    D_\beta \longrightarrow \bullet \longrightarrow M_\alpha &~~~~~~~~~~
                -\Gamma_{M_\alpha, D_\beta}^T,\cr
    D_\beta \longrightarrow \bullet \longleftarrow M_\alpha &~~~~~~~~~~
                -\Gamma_{M_\alpha, D_\beta}^TC^{-1},\cr
    M_\beta \longrightarrow \bullet \longrightarrow D_\alpha &~~~~~~~~~~
                -\Gamma_{D_\alpha, M_\beta}^T,\cr
    M_\beta \longleftarrow \bullet \longrightarrow D_\alpha &~~~~~~~~~~
                -C^T\Gamma_{D_\alpha, M_\beta}^T.\cr
}\eqno(29)
$$\par
   Next, we consider the Majorana - Majorana vertices.  For the lagrangian,
(27), we assign the vertex functions as follows:
$$
\eqalignno{
    M_\alpha \longleftarrow \bullet \longleftarrow M_\beta &~~~~~~~~~~
                \Gamma_{M_\alpha, M_\beta},           & (30)  \cr
    M_\alpha \longrightarrow \bullet \longleftarrow M_\beta &~~~~~~~~~~
                (C^{-1})^T\Gamma_{M_\alpha, M_\beta}, &  (31) \cr
    M_\alpha \longleftarrow \bullet \longrightarrow M_\beta &~~~~~~~~~~
                \Gamma_{M_\alpha, M_\beta}C,          &  (32) \cr
    M_\beta \longrightarrow \bullet \longrightarrow M_\alpha &~~~~~~~~~~
                -\Gamma_{M_\alpha, M_\beta}^T,        &  (33) \cr
    M_\beta \longrightarrow \bullet \longleftarrow M_\alpha &~~~~~~~~~~
                -\Gamma_{M_\alpha, M_\beta}^TC^{-1},  &  (34) \cr
    M_\beta \longleftarrow \bullet \longrightarrow M_\alpha &~~~~~~~~~~
                -C^T\Gamma_{M_\alpha, M_\beta}^T.     &  (35) \cr
}$$
Note that these Grace line rules are valid under the condition, 
$\alpha \ge \beta$.  When the Majorana spinor species at the vertex
has an opposite order, rewriting the lagrangian (27) as
$$
   {\cal L} = \sum_{\alpha\ge\beta}
              \bar\Psi(M_\beta)\Gamma^c_{\beta,\alpha}\Psi(M_\alpha).
\eqno(36)
$$
where
$$
   \Gamma^c_{\beta,\alpha}= C\Gamma^T_{\alpha,\beta}C^{-1}.
\eqno(37)
$$
we obtain the following rule;
$$
\eqalignno{
    M_\beta \longleftarrow \bullet \longleftarrow M_\alpha &~~~~~~~~~~
                C\Gamma^T_{M_\alpha, M_\beta}C^{-1},           & (38)  \cr
    M_\beta \longrightarrow \bullet \longleftarrow M_\alpha &~~~~~~~~~~
                -\Gamma^T_{M_\alpha, M_\beta}C^{-1}, &  (39) \cr
    M_\beta \longleftarrow \bullet \longrightarrow M_\alpha &~~~~~~~~~~
                C\Gamma^T_{M_\alpha, M_\beta},          &  (40) \cr
    M_\alpha \longrightarrow \bullet \longrightarrow M_\beta &~~~~~~~~~~
                -C^{-1}\Gamma_{M_\alpha, M_\beta}C,        &  (41) \cr
    M_\alpha \longrightarrow \bullet \longleftarrow M_\beta &~~~~~~~~~~
                -C^{-1}\Gamma_{M_\alpha, M_\beta},  &  (42) \cr
    M_\alpha \longleftarrow \bullet \longrightarrow M_\beta &~~~~~~~~~~
                \Gamma_{M_\alpha, M_\beta}C.     &  (43) \cr
}$$
Note that (31)=(42), (32)=(43), (34)=(39) and (35)=(40).
These equalities provide us one of the  
consistency check of the Grace line rule, since  the rules (31)$-$(35)  
are obtained from (30) while (39)$-$(43) are obtained from (38). 
\medskip

 From (28), (29) and above equations, one finds the following general
rule of how and where to put the charge conjugation matrix $C$:\par
\noindent{$\underline{\bf Rule~ 3}$}\par
$$
\eqalign{
     \longleftarrow \bullet ---~~ =& -C(\longrightarrow\bullet ---), \cr 
     \longrightarrow \bullet ---~~ =& -C^{-1}(\longleftarrow\bullet ---), \cr
     ---\bullet\longleftarrow~~ =& ~~~~~~(---\bullet\longrightarrow)C^{-1}, \cr
     ---\bullet\longrightarrow~~ =& ~~~~~~(---\bullet\longleftarrow)C,
}$$
where $---$ is a fermion line whose F-direction can be arbitrary.
\medskip
     
     Finally, a comment is in order concerning the fermion number violating
chargino vertex.  MSSM contains two charginos $\tilde\chi^\pm_1$ and 
$\tilde\chi^\pm_2$. Remember that we have defined the {\it positive} chargino 
as Dirac fermion. Their interactions are given by
$$
\eqalign{
     {\cal L}_{CD} =& \bar\Psi(\tilde\chi^-_i)\Gamma_{C,e}\Psi(e) 
                   +\bar\Psi(\tilde\chi^+_i)\Gamma_{C,\nu}\Psi(\nu)+h.c., \cr
     {\cal L}_{CM} =& \bar\Psi(\tilde\chi^+_i)\Gamma_{C_i,M_\alpha}
                     \Psi(M_\alpha) + h.c. 
}\eqno(44)
$$
Denoting 
$$
   \Gamma_{e,C} = \gamma_0\Gamma_{C,e}^\dagger \gamma_0,~~~
   \Gamma_{\nu,C} = \gamma_0\Gamma_{C,\nu}^\dagger \gamma_0,~~~
   \Gamma_{M_\alpha,C_i} = \gamma_0\Gamma_{C_i,M_\alpha}^\dagger \gamma_0,
\eqno(45)
$$
the Grace line is represented by
$$
\eqalign{
  \tilde\chi^+_i ~~ \longleftarrow \bullet\longleftarrow ~~\nu ~~~~~~~~~~~~
   ~~~&~~~~~~~~\Gamma_{C,\nu}      \cr
  \nu ~~ \longleftarrow \bullet\longleftarrow ~~\tilde\chi^+_i ~~~~~~~~~~~~~
   &~~~~~~~~\Gamma_{\nu,C}      \cr
   \tilde\chi^+_i ~~ \longrightarrow \bullet\longleftarrow ~~e^- ~~~~~~~~~~~~~
   &~~~~~~~~-C\Gamma_{C,e}        \cr
    e^- ~~ \longleftarrow \bullet\longrightarrow ~~\tilde\chi^+_i ~~~~~~~~~~~~~
   &~~~~~~~~\Gamma_{e,C}C        \cr
  \tilde\chi^+_i~~ \longleftarrow \bullet\longleftarrow ~~M_\alpha~~~~~~~~~~~~
   &~~~~~~~~\Gamma_{C_i,M_\alpha}      \cr
   M_\alpha~~ \longleftarrow \bullet\longleftarrow ~~\tilde\chi^+_i~~~~~~~~~~~~
   ~&~~~~~~~~\Gamma_{M_\alpha,C_i}      \cr
}\eqno(46)
$$
The first two Grace line rules are the same as those for the conventional 
Dirac fermion 
vertex, while the subsequent two lines are new, violating the fermion 
number conservation.\par
    Note that the charge conjugation matrix $C$ appear explicitly in the
Grace line rule (46).  The way $C$ appears in (46) obeys the general 
Grace line rule, {\bf rule~3}, although the vertices corresponding to 
$\Gamma_{C,e}$ and $\Gamma_{e,C}$ do not exist in Grace line amplitudes.
\medskip
     
     In case the {\it negative} charginos are defined as Dirac fermions 
(as is preferred by some authors), rewriting the lagrangian 
(44) as follows,
$$
\eqalign{
{\cal L}_{CD}=&\bar\Psi(e^+)\Gamma^c_{e,C}\Psi(\tilde\chi^+_i)
              +\bar\Psi(\bar\nu)\Gamma^c_{\nu,C}\Psi(\tilde\chi^-_i) + hc,\cr
{\cal L}_{CM}=&\bar\Psi(M_\alpha)\Gamma^c_{M_\alpha,C_i}
\Psi(\tilde\chi^-_i)+hc,
}\eqno(47)
$$
where
$$
    \Gamma^c_{A,B} \equiv C\Gamma^T_{B,A}C^{-1},   ~~~~~
    (A,B) = (e,C),~~(\nu,C),~~(M_\alpha,C_i),
\eqno(48)
$$
one obtains the corresponding Grace line rule:
$$
\eqalign{
  \nu ~~ \longrightarrow \bullet\longleftarrow ~~\tilde\chi^-_i ~~~~~~~~~~~~
   &~~~~~~~~-C^{-1}\Gamma^c_{\nu,C}      \cr
  \tilde\chi^-_i ~~ \longleftarrow \bullet\longrightarrow ~~\nu ~~~~~~~~~~~~
   ~~&~~~~~~~~\Gamma^c_{C,\nu}C      \cr
   e^- ~~ \longrightarrow \bullet\longrightarrow ~~\tilde\chi^+_i ~~~~~~~~~~~~
   &~~~~~~~~\Gamma^c_{e,C}        \cr
   \tilde\chi^-_i ~~ \longrightarrow \bullet\longrightarrow ~~e^- ~~~~~~~~~~~~
   &~~~~~~~~\Gamma^c_{C,e}        \cr
  M_\alpha~~ \longleftarrow \bullet\longleftarrow ~~\tilde\chi^-_i~~~~~~~~~~~~
   &~~~~~~~~\Gamma^c_{M_\alpha,C_i}      \cr
  \tilde\chi^-_i~~ \longleftarrow \bullet\longleftarrow ~~M_\alpha~~~~~~~~~~~
   &~~~~~~~~\Gamma^c_{C_i,M_\alpha}      \cr
}\eqno(49)
$$
The second term of ${\cal L}_{CD}$ now violates the fermion number 
conservation, 
which can be explicitly understood from (49).  \par
   The rule (49) is equivalent to the rule (46) and therefore, 
(49) is derived from (46) without referring to the lagrangian (47).
For example, the Grace line rule of the second line of (49) is derived
from (46) as follows,
$$
   \tilde\chi^-_i ~~ \longleftarrow \bullet\longrightarrow ~~\nu ~~
   = -C(
   \tilde\chi^+_i ~~ \longrightarrow \bullet\longrightarrow ~~\nu ~~)
   = -C(-\Gamma^T_{\nu,C})
   = \Gamma^c_{C,\nu}C,
\eqno(50)
$$
where {\bf rule~2}, {\bf rule~3} and (37) are used.\par
   Note that the expression 
of the Grace line amplitudes is independent of the convention of whether one 
assigns the fermion number $+1$ for {\it positive} charginos or {\it negative}
charginos.  The differences at the vertices, (46) vs. (49), 
are compensated by the chargino propagators and external spinors.
\vskip 2 truecm

\noindent{\bf 4. Identical Particles}\par
     For the vertex consisting of two identical Majorana particles ($\alpha
=\beta$ in (27) and (30)$-$(43)), twelve Grace line rules in (30)$-$(43) are not
independent from each other, but the first six  rules are identical to
the second six ones.  Therefore, the expression of the first six vertices 
is found among  the second six ones, 
namely, (30)=(38), (31)=(39),  etc.  Consequently, the Majorana-Majorana
vertex function $\Gamma_{\alpha,\alpha}$ must
satisfy the following condition,
$$
     \Gamma_{\alpha,\alpha} = C\Gamma_{\alpha,\alpha}^TC^{-1},
        ~~~~~~\alpha {\rm ~not~ summed}.
\eqno(51)
$$
This is equivalent to saying  that there are no vector current nor 
tensor current in the Majorana bilinear form, a well known 
property of Majorana particles, as
$$
   \Gamma=(1,\gamma_5, \gamma_\mu, \gamma_\mu\gamma_5, \sigma_{\mu\nu})
     ~~~~~\to ~~~~~
   \Gamma^c\equiv C\Gamma C^{-1}=
          (1,\gamma_5, -\gamma_\mu, \gamma_\mu\gamma_5, -\sigma_{\mu\nu}).
\eqno(52)
$$
See also the item 2 of Appendix C.  In fact, the MSSM 
lagrangian ${\cal L}_{MM}$
has a following form for the diagonal Majorana-Majorana interaction part
(see e.g. [8]),
$$
    {\cal L}_{MM} = c_Z\bar\Psi_M \gamma_\mu \gamma_5Z^\mu 
                                               \Psi_M
                  + c_H\bar\Psi_M (1 + b_H\gamma_5) H \Psi_M
                  + c_G\bar\Psi_M (1 + b_G\gamma_5) G \Psi_M,
\eqno(53)
$$
where $H$ represents any of the three neutral Higgs particles and $G$
is a neutral Goldstone boson.  
In addition, we have to multiply the vertex functions
by a statistical factor 2.
\vskip 2 truecm

\noindent{\bf 5 Example}\par
     In this section let us consider two examples in which F-direction
is not conserved on a Grace line.  The first example is the
Majorana-Dirac-fermion annihilation process with two vertices and one
Majorana propagator,
$$
 <0\vert {\cal L}_{MM}{\cal L}_{MD}c_m^\dagger a_n^\dagger \vert 0> = 
          -\bar v_m \left\lgroup\matrix
          {\Gamma_{m,p}\cr\Gamma^c_{m,p}\cr}\right\rgroup S_p\Gamma_{p,n} u_n,
\eqno(54)
$$
where $p$ stands for the Majorana species of the propagator $S$ and 
two vertex functions correspond to the case  $m\ge p$ and $m\le p$, 
respectively.
There are four possible ways of assigning the F-direction on the Majorana
propagator.  
$$
\eqalign{
   \tilde\chi^0_m~  
   \longrightarrow\bullet\longleftarrow\longleftarrow\bullet\longleftarrow 
   ~D_n~~~ &~~~~~  \to~~~
    (-)u_m^T\left\lgroup\matrix{
      (C^{-1})^T\Gamma_{m,p}\cr -\Gamma^T_{p,m}C^{-1}}\right\rgroup
                     S_p(k)\Gamma_{p,n}u_n ,\cr
    p_m\rightarrow~~~~~\leftarrow k~~~~~~\leftarrow p_n ~~~~~~~    &\cr
   \tilde\chi^0_m~  
   \longrightarrow\bullet\longrightarrow\longrightarrow\bullet\longleftarrow 
   ~D_n~~~&~~~~   \to~~~
    (-)u_m^T\left\lgroup\matrix{
    -C^{-1}\Gamma_{m,p}C\cr -\Gamma^T_{p,m}}\right\rgroup
                   (-S^T_p(-k))(C^{-1})^T\Gamma_{p,n}u_n. \cr 
    p_m\rightarrow~~~~~\leftarrow k~~~~~~\leftarrow p_n~~~~~~~     &\cr
   \tilde\chi^0_m~  
   \longleftarrow\bullet\longleftarrow\longleftarrow\bullet\longleftarrow 
   ~D_n~~~ &~~~~~  \to~~~
    (-)\bar v_m\left\lgroup\matrix{
      \Gamma_{m,p}\cr C\Gamma^T_{p,m}C^{-1}}\right\rgroup
                     S_p(k)\Gamma_{p,n}u_n ,\cr
    p_m\rightarrow~~~~~\leftarrow k~~~~~~\leftarrow p_n~~~~~~~     &\cr
   \tilde\chi^0_m~  
   \longleftarrow\bullet\longrightarrow\longrightarrow\bullet\longleftarrow 
   ~D_n~~~&~~~~   \to~~~
    (-)\bar v_m\left\lgroup\matrix{
          \Gamma_{m,p}C\cr C\Gamma^T_{p,m}}\right\rgroup
                   (-S^T_p(-k))(C^{-1})^T\Gamma_{p,n}u_n. \cr 
    p_m\rightarrow~~~~~\leftarrow k~~~~~~\leftarrow p_n ~~~~~~~    &\cr
}\eqno(55)
$$
Upon using (21), (37) and 
$$
    -CS(-k)^T(C^{-1})^T = CS(-k)^T C^{-1}=
         C {{-\slash k^T+m}\over{k^2-m^2+i\epsilon}} C^{-1} = S(k),
\eqno(56)
$$
it is easy to prove that the four expressions of (55) are indeed 
identical to (54).  Our convention of assigning the M-direction 
and F-direction
to  Majorana particles chooses the first assignment of (55).
\medskip    
 
     The next example, $\tilde\chi^-_i \rightarrow e^- \tilde {\bar\nu}_L$, 
contains the fermion number violating chargino interaction. 
The lagrangian is given by (44).  The Grace 
line for this process is depicted as follows,
$$
\eqalign{ 
 e^-~~~ \longleftarrow\bullet&\longrightarrow \tilde\chi^+_i \cr
  p_m\leftarrow~~~&~~~\leftarrow p_n
}\eqno(57)
$$
 From (46) and Table 2, one finds
$$
 {\cal M} =  \bar u(p_m)(\Gamma_{e,C}C) \bar v^T(p_n) 
          =   \bar u(p_m)\Gamma_{e,C} u(p_n).
\eqno(58)
$$
This can be explicitly evaluated as follows,
$$
\eqalign{
   <\tilde {\bar\nu}_L e^-\vert {\cal L} \vert \tilde\chi^-_i> 
   =&  <\tilde {\bar\nu}_L e^- \vert 
    \bar\Psi(e)\Gamma_{e,C}\Psi(\tilde\chi^-_i) \vert \tilde\chi^-_i> \cr
  = & <\tilde {\bar\nu}_L \vert a_e
    \bar\Psi(e)\Gamma_{e,C}\Psi^c(\tilde\chi^+_i) b^\dagger_{\chi}
                   \vert 0> \cr
  =& \bar u(e)\Gamma_{e,C} u(\tilde\chi_i)
}\eqno(59)
$$
At first sight it seems odd that spinor $u(\tilde \chi_i)$ appears in the
amplitude in spite of the antifermion in the initial state.  We should 
recall, however, that whether chargino behaves like fermion or antifermion 
is not determined
by the convention we adopt but by the nature of the interaction vertex.

\vskip 1.5 truecm
\vfill\eject

\noindent{\bf 6. Comparison with Denner's method}\par
    Denner et al.[9] have introduced the lagrangian 
written in terms of  the charge conjugated state
in order to eliminate the charge conjugation matrix $C$ from the 
Feynman amplitudes.  Although for our purpose we don't find any particular 
advantage in adopting 
their method, it is instructive to compare our method with theirs.  The
concept of orientation (fermion flow) which they have  introduced 
in order to
specify whether the conventional vertices are used or the charge conjugated
ones are used corresponds to our two ways of 
displaying Grace line.  Although as our convention we have decided 
to stick to $G_{m,n}$ with 
$m>n$, we can equally 
use the Grace line $G_{n,m}$ with $m>n$.
The main point of ref.[9] is that all the charge
conjugation matrices appearing in the Grace line amplitudes can be combined
with the vertex function $\Gamma$  to form the charge conjugated one 
$\Gamma^c$ and can be eliminated completely.  \par 
     Let us prove that two methods are equivalent by showing that the 
Grace line amplitude can be brought in the form that is obtained 
from the prescription given by Denner et al.   This is done by showing
(1) charge conjugation matrices, transposed propagators and transposed
vertices  can fully be eliminated in favour of $\Gamma^c=C\Gamma^T C^{-1}$ and
$S(k) = CS^T(-k)C^{-1}$ (see (56)).  (2)The resulting expression contains
$\Gamma^c$ at the proper vertices.  (3) The overall sign of {\it the Feynman
amplitude} \footnote{*}{Here we refer to the Feynman amplitude and not
the Grace line amplitude, since the sign of the Grace line amplitude 
$[G_{p,q}]$ is not independent from the partition (4) and therefore 
from $sign(P)$.} 
is given by $(-)^{P+V}$  when expressed without charge 
conjugation matrix $C$. Here $P$ is due to the permutation of spinors 
and $V$ is the number of $v$ and $\bar v$ in the Feynman amplitude.\par
     First we note that as our convention the M-direction of the internal 
spinors in a Grace line is, irrespective of Majorana or Dirac 
particle,  always taken in the A-direction and that 
the F-direction of the internal Majorana particles 
on the Grace line is also taken in the 
A-direction.   Consequently, there are only two sources of the charge 
conjugation matrix; from the external spinor wave functions and 
from the clashing vertices ( vertices where two F-directions are not
same).  
 From the propagator we don't get any $C$.   With the use of (21), we
rewrite Table 2 in the form which has no transposed wave functions, which 
is shown in Table 3.  Note that the minus signs coming from the {\bf rule 1}
are now included in Table 3. \par
 
\topinsert
{\offinterlineskip \tabskip=0pt
\halign{ \strut
  \vrule#& \quad # \quad &
  \vrule#& \quad # \quad &
  \vrule#& \quad # \quad &
  \vrule#& \quad # \quad &
  \vrule#& \quad # \quad &
  \vrule#  \cr
\noalign{\hrule}
  & F-direction  && $\longleftarrow\bullet$ &&  $\longrightarrow\bullet$ &&
                    $\bullet\longleftarrow$ &&  $\bullet\longrightarrow$ & \cr
  &  M-direction &&                 &&               && &&              & \cr
\noalign{\hrule}
  & $\longleftarrow$  &&   $~\bar u(p)$   &&    $ -\bar u(p)C$  
                      &&   $~u(p)$        &&    $ ~C^{-1}u(p)$ & \cr
  & $\longrightarrow$ &&   $-\bar v(p)$   &&    $ ~\bar v(p)C$  
                      &&   $-v(p)$        &&    $ -C^{-1}v(p)$ & \cr
\noalign{\hrule} }} 
\medskip
\noindent{ Table 3}~~~Spinor assignment at left endpoint and 
at right endpoint.
\endinsert

 From Table 3 we notice that the charge conjugation matrix appears 
in connection with external spinors whose F-direction is opposite to 
the A-direction.\par
       At vertices, the charge conjugation matrix appears  
when two F-directions clash.  From {\bf rule 3} one finds, 
$$
  p~~\leftarrow\bullet\rightarrow ~~q~~~ =~~~~~ \hat \Gamma C
~~~~~~with~~~~~~
     \hat\Gamma = \cases{\Gamma_{D,M}~~~~~\mwhen~  (D,M)\cr
                  \Gamma^c_{M,D}~~~~~\mwhen~  (M,D)\cr
                  \Gamma_{\alpha,\beta} ~~~~~~   \mwhen~ (M_\alpha,M_\beta)\cr
                  \Gamma^c_{\beta,\alpha}~~~~~~ \mwhen~ (M_\beta,M_\alpha)\cr
                  \Gamma_{e,C_i}~~~~~\mwhen~(e, C_i)\cr},
\eqno(60)
$$
$$
  p~~\rightarrow\bullet\leftarrow ~~q~~~ =~~~~~~ -C^{-1} \hat \Gamma 
~~~~~~with~~~~~~
     \hat\Gamma = \cases{\Gamma_{M,D}~~~~~\mwhen~  (M,D)\cr
                  \Gamma^c_{D,M}~~~~~\mwhen~  (D,M)\cr
                  \Gamma_{\alpha,\beta}~~~~~~    \mwhen~ (M_\alpha,M_\beta)\cr
                  \Gamma^c_{\beta,\alpha}~~~~~~ \mwhen~ (M_\beta,M_\alpha)\cr
                  \Gamma_{C_i,e}~~~~~\mwhen~(C_i,e)\cr}.
\eqno(61)
$$
Here the charge conjugated vertex function is defined by
$$
   \Gamma^c_{p,q} = C \Gamma^T_{q,p}C^{-1},
\eqno(37)=(62)
$$
and the vertex functions, $\Gamma_{MD}$, $\Gamma_{\alpha\beta}$ etc
are defined by our lagrangian (24), (27) and (44), namely,
$$
  \Gamma_{A,B} =  ~~~~~A~\leftarrow\bullet\leftarrow~B.
\eqno(63)
$$
The rules (60) and (61) are summarized as follows: $C$ appears to the
right of $\hat\Gamma$ when two F-directions depart from the vertex,
and $-C^{-1}$ appears to the left of $\hat\Gamma$ when two
F-direction clash at the vertex.  $\Gamma^c$ is used when the
"reversed" vertex appears.  Note that 
the word "reversed" is understood in the sense
that the order of the
particle species (in the case of Dirac particles also their
F-direction) is reversed compared with the order appearing in the 
model lagrangian. Therefore, it is important to fix the lagrangian
we use for the definition of $\Gamma_{AB}$.  In particular, for the
chargino-Dirac fermion interactions, which violate fermion
number conservation, we have to decide which of the two equivalent
forms, namely, (44) or (47), we adopt for our lagrangian.  If we 
had adopted the lagrangian (47) instead of (44), then 
$\Gamma^c$ in (47) would have been considered as $\Gamma$   that is used in
(60) and (61).  See the footnote between {\it rule 1} and {\it rule
2}.\par
 
     In order to see where the charge conjugation matrices
appear and how they are eliminated in the Grace line in which the F-direction
is not conserved at some vertices, it is sufficient to consider the following 
three spinor chains, each of which is a  section of a certain Grace line,
$$
\eqalignno{
   \cdots\leftarrow\bullet&\rightarrow\rightarrow\bullet\bullet\bullet\bullet
      \bullet\rightarrow\rightarrow\bullet\leftarrow\cdots   &  (64a) \cr
      \rightarrow\bullet&\rightarrow\rightarrow\bullet\bullet\bullet\bullet
   \bullet\rightarrow\rightarrow\bullet\leftarrow\cdots  & (64b)\cr
      \cdots\leftarrow\bullet&\rightarrow\rightarrow\bullet\bullet
   \bullet\bullet\bullet \rightarrow\rightarrow\bullet\rightarrow~~. &(64c)\cr   
}$$
Here $\cdots$ means that there exist propagators to the left or 
to the right of the chain,  
namely the external spinors are not involved in this sequence of spinors 
and $\bullet\bullet\bullet\bullet\bullet$ stands for the vertices 
and propagators
with F-direction opposite to the A-direction.  The spinor chain of the
type (64a) appears somewhere inside the Grace line.  When the F-directions
of the spinors at both ends of the Grace line are  in the A-direction,
this kind of the spinor chain is the only possible source of the charge 
conjugation matrix.
If the F-direction of the left endpoint is opposite to the  A-direction, 
the spinor chain of the type (64b) produces charge conjugation matrices.
while if the F-direction of the right endpoint is opposite to the A-direction,
the spinor chain of the type (64c) produces charge conjugation matrices.
The shortest chain of the type (64a) is the chain with one propagator and 
that of the type (64b) and (64c)  is the chain without propagators.\par 
     The Grace line amplitude corresponding to (64) is given by
$$
\eqalignno{
     \cdots\hat\Gamma C (-S^T)(-\Gamma^T)\cdots(-S^T)(-C^{-1})\hat\Gamma....,
     &=\cdots\hat\Gamma S \Gamma^c\cdots S \hat\Gamma         &  (65a)\cr
     \left\lgroup \matrix {-\bar u \cr \bar v}\right\rgroup
        C (-\Gamma^T)(-S^T)\cdots (-S^T)(-C^{-1}\hat\Gamma)\cdots
     &=\left\lgroup \matrix {\bar u \cr -\bar v}\right\rgroup
        \Gamma^c S\cdots S\hat\Gamma\cdots                         & (65b) \cr
     \cdots\hat\Gamma C (-S^T)\cdots(-S^T)(-\Gamma^T)C^{-1}
     \left\lgroup\matrix{ u \cr -v\cr}\right\rgroup
         &= \cdots\hat \Gamma S\cdots S \Gamma^c 
          \left\lgroup\matrix{ u \cr -v\cr}\right\rgroup        &  (65c) \cr
}$$           
   (65) proves that charge conjugation matrices and transposed vertices as
well as the transposed propagators have completely disappeared  from the 
Grace line amplitude in favour of the appearance of $\Gamma^c$.  In addition,
(65) also proves that the sign of the Grace line amplitude is given by 
$(-)^{Vg}$, where $V_g$ is the total number of $v$ and $\bar v$ appearing in a
Grace line.  When combined with the sign coming from the permutation of 
spinors,
one finds that the sign of {\it the amplitude} is given\footnote{*} 
{  At first sight, the value of $V=\sum V_g$ seems not unique
since the external Majorana can be taken as particle as well as antiparticle.
Note, however,  that the assignment of the spinor wave 
functions is unique, provided one consistently uses Table 2.
As shown in Table 2, when the external particle has the 
M-direction which is opposite to the A-direction, one has to use $v$ or
$\bar v$, depending on whether the external particle is at the right 
endpoint or at the left endpoint, but independent of the F-direction.  
Since the assignment of $u$ and $v$  to each external spinor is unique, 
the value
$V$, the total number of $v$ and $\bar v$ appearing in {\it the Feynman 
amplitude} is also unique.}
by $msign(P)(-)^V$ with $V=\sum V_g$,
in accordance with Denner et al.  Further, (60), (61) and (65) tell us that 
we should 
use the $\Gamma^c$ at the clashing vertex which has a reserved order of 
spinors compared to the vertex which defines $\Gamma$ and at the non-clashing
vertex when the F-direction is opposite to the A-direction.  
This completes the proof of the equivalence
of the two methods. \par
\medskip
 
     As an illustration, we compare the two methods in two examples.  
We start with the decay of a scalar ($\Phi$) into  a Dirac ($D$) 
and a Majorana ($M$) particle, an example discussed by Denner et al.[9].
Suppose that the underlying lagrangian is given by
$$
     {\cal L} = \bar \Psi(D)\Gamma_{D,M} \Psi(M)\Phi + 
                \bar \Psi(M)\Gamma_{M,D} \Psi(D)\Phi^*.
\eqno(66)
$$
Assign number 1 to the Dirac particle and number 2 to  the Majorana 
particle.  There
is only one pair $(2,1)$ and the Grace line and the corresponding Grace 
line amplitude $[G_{2,1}]$ is given by
$$
\eqalign{
  M \longleftarrow\bullet\longrightarrow  D~~~~~~~~~\to&   
  ~~~~~[G_{2,1}] = (-)\bar u(p_m) [-C^T\Gamma^T_{D,M}]  (\bar u(p_d))^T \cr
   ~~ p_m\leftarrow~~~~~\rightarrow p_d ~~~~~~~~~~~~~~ &    
  ~~~~~~~~~~~~ =- \bar u(p_m)\Gamma^c_{D,M} v(p_d). \cr
}\eqno(67)
$$
This corresponds to the case of Fig.4b of ref.[9].\footnote{**}
{It seems there is a misprint in (3.1) of ref.[9].  For
the decay of $\Phi$, one should use the third term of the second line of 
(2.1).  Then, Denner's $k^{i*}\Gamma_i$ corresponds to our $\Gamma_{D,M}$ 
and $k^{i*}\Gamma^\prime_i$ to $\Gamma^c_{D,M}$.}
The difference of the 
overall sign is irrelevant.  The amplitude corresponding to Fig.4c of ref.[9],
whose orientation is from Majorana to Dirac,  corresponds to $[G_{1,2}]$.
The explicit construction based on our Grace line rule gives
$$
\eqalign{
  D \longleftarrow\bullet\longrightarrow  M~~~~~~~~~\to &   
  ~~~~~[G_{1,2}] = (-)\bar u(p_d) \Gamma_{D,M} C (\bar u(p_m))^T \cr
   ~~ p_d\leftarrow~~~~~\rightarrow p_m ~~~~~~~~~~~~~~ &    
  ~~~~~~~~~~~~ = -\bar u(p_d)\Gamma_{D,M} v(p_m), \cr
}\eqno(68)
$$
which indeed coincides with the result given by Denner et al., and which
also proves that $[G_{1,2}]=-[G_{2,1}]$.\par

   The next example discussed by Denner et al.[9] is  the scattering $D_a D_b 
\to \Phi_c\Phi_d$, where it is assumed that $D_a$ and $D_b$ are not 
identical particles.  There is only one Feynman diagram which has a
Majorana propagator, but there are two possible orientations as shown
in Figure 2.

\centerline{\psfig{file=fig2.epsf}}
\sevenrm
\baselineskip = 8pt
\centerline{
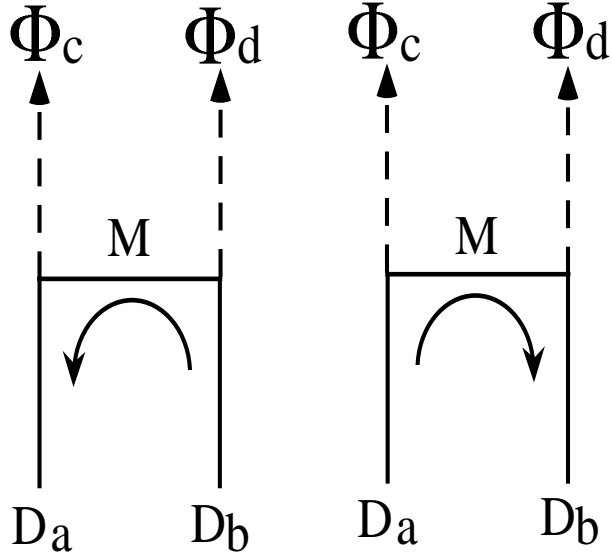
\noindent{Figure 2.}  Two possible orientations for the process
$D_aD_b\to \Phi_c\Phi_d$ as studied  in ref.[9]. }
\tenrm
\baselineskip = 14pt
\vskip 1 true cm 

Corresponding to (3.3a) and (3.3b) of ref.[9], we have
$$
\eqalignno{
   \longrightarrow\bullet\longleftarrow\bullet\longleftarrow~~~~~~~~\to&
     ~~~~~~~~  [G_{a,b}]=(-)u^T_a(-\Gamma^T_{M,D}C^{-1})S(p_b-p_d)
                      \Gamma_{M,D}u_b   \cr
     p_a \rightarrow~~~~~~~~\leftarrow p_b~~~~~~~~~~&
    ~~~~~~~~~~~~~~~=-\bar v_a \Gamma^c_{M,D}S(p_b-p_d)\Gamma_{M,D}u_b, &(69)\cr
   \longrightarrow\bullet\longleftarrow\bullet\longleftarrow~~~~~~~~\to&
     ~~~~~~~~  [G_{b,a}]=(-)u^T_b(-\Gamma^T_{M,D}C^{-1})S(p_a-p_c)
                      \Gamma_{M,D}u_a   \cr
     p_b \rightarrow~~~~~~~~\leftarrow p_a~~~~~~~~~~&
    ~~~~~~~~~~~~~~~=-\bar v_b \Gamma^c_{M,D}S(p_a-p_c)\Gamma_{M,D}u_a,&(70) \cr
} $$
respectively.   Note that $p_a+p_b=p_c+p_d$.
\vskip 1 truecm

\noindent{\bf Acknowledgement}\par
    I would like to thank Prof. T. Kaneko and Prof. H. Tanaka (Rikkyo 
University) for fruitful discussions and an efficient introduction of
{\tt GRACE} system.

\vskip 2 truecm

\noindent{\bf References}\par  
\item{[1]} T. Kaneko, in New computing techniques in physics research,
eds., D. Perret-Gallix and W. Woj\-cik, (1990) p.555.
\item{[2]} T. Kaneko and H. Tanaka, Proc. in the second workshop on JLC, ed.
S. Kawabata, (1991) p.250.
\item{[3]} E.F. Boos et al., New computing techniques in physics research,
vol.1, eds., D. Perret-Gallix and W. Woj\-cik, (1990) p.573, 
{\it ibid} vol.2, eds., D. Perret-Gallix et al., (1992) p.665, 
A.E. Pukhov, {\it ibid} vol.4, ed., K.-H.
Becks, D. Perret-Gallix (1993) p.473.
\item{[4]} R. Mertig et al.,  Compt. Phys. Commun. 64 (1991) 345.
\item{[5]} T. Ishikawa et al., KEK preprint 92-19 (1993) The Grace
manual, ver.1.0.
\item{[6]} J. Fujimoto et al., Compt. Phys. Commun. 100(1997) 128.
\item{[7]} Y. Kurihara et al., Phys.Lett.B349(1995) 367.
\item{[8]} H.E. Haber and G.L. Kane, Phys. Rep. 117(1985) 75.
\item{[9]} A. Denner et al., Nucl.Phys. B387(1992) 467.

\vfill\eject

\noindent{\bf Appendix A. Charge conjugation and Majorana particles}\par
   Dirac spinor field operators are expanded in terms of the creation and 
annihilation operators:
$$
\eqalign{
    \Psi(x) =& \sum_\lambda \int{{d^3p}\over{(2\pi)^3 2p_0}}
    [a(p\lambda)u(p\lambda)e^{-ipx}+b^\dagger(p\lambda)v(p\lambda)e^{ipx}],\cr
 \bar\Psi(x) =& \sum_\lambda \int{{d^3p}\over{(2\pi)^3 2p_0}}
    [b(p\lambda)\bar v(p\lambda)e^{-ipx}+a^\dagger(p\lambda)\bar u(p\lambda)
        e^{ipx}],\cr
}\eqno(A.1)
$$\par
     The unitary operator ${\cal C}$, called as charge conjugation 
operator, is defined by the following properties           
$$
\eqalign{
  {\cal C} a^\dagger(p,\lambda)\vert 0>
   =& \epsilon^* b^\dagger(p,\lambda)\vert 0>,\cr
  {\cal C}^2 a^\dagger(p,\lambda)\vert 0>
   =& a^\dagger(p,\lambda)\vert 0>,\cr
   {\cal C}\vert 0> =& \vert 0>.
}\eqno(A.2)
$$ 
 From these three conditions, it follows
$$
\eqalign{
       {\cal C} a^\dagger(p,\lambda){\cal C}^{-1} 
       =& \epsilon^* b^\dagger(p,\lambda),\cr    
       {\cal C} b^\dagger(p,\lambda){\cal C}^{-1} 
       =& \epsilon a^\dagger(p,\lambda),\cr    
       {\cal C} a(p,\lambda){\cal C}^{-1} 
       =& \epsilon b(p,\lambda),\cr    
       {\cal C} b(p,\lambda){\cal C}^{-1} 
       =& \epsilon^* a(p,\lambda),\cr    
       {\cal C} \Psi(x){\cal C}^{-1} 
        =& \epsilon \Psi^c(x),\cr
       {\cal C} \bar\Psi(x){\cal C}^{-1} 
       =& \epsilon^* \overline{\Psi^c(x)},\cr    
}\eqno(A.3)
 $$
where 
$$
\eqalign{
    \Psi(x)^c =& \sum_\lambda \int{{d^3p}\over{(2\pi)^3 2p_0}}
    [b(p\lambda)u(p\lambda)e^{-ipx}+a^\dagger(p\lambda)v(p\lambda)e^{ipx}],\cr
 \bar\Psi(x)^c =& \sum_\lambda \int{{d^3p}\over{(2\pi)^3 2p_0}}
    [a(p\lambda)\bar v(p\lambda)e^{-ipx}+b^\dagger(p\lambda)\bar u(p\lambda)
        e^{ipx}].\cr
}\eqno(A.4)
$$
Using the four-by-four unitary matrix $C$ (charge conjugation matrix), which
satisfies the conditions,
$$
     C\bar u(p,\lambda)^T = v(p,\lambda),~~~~~~~
     C\bar v(p,\lambda)^T = u(p,\lambda),
\eqno(A.5)
$$
$\Psi(x)$ and its charge conjugation partner $\Psi^c(x)$ are related as 
follows;
$$
      \Psi^c= C\bar\Psi^T,~~~~~~~\overline{\Psi^c} = \Psi^T(C^T)^{-1},
\eqno(A.6)
$$
and 
$$
\eqalign{
  {\cal C}\Psi(x)^T {\cal C} =& \epsilon \bar\Psi(x) C^T 
                             = \epsilon \Psi^{cT}(x), \cr
  {\cal C}\bar\Psi^T(x){\cal C} =&\epsilon^*C^{-1} \Psi(x)
                                =\epsilon^*\overline{\Psi^c}^T(x).
}\eqno(A.7)
$$
The charge conjugation matrix $C$ is subject to several conditions. 
First, from the normalization condition, $\bar u u = - \bar v v = 1$ 
and (A.5), one finds that $C$ is antisymmetric,
$$
     C^T = -C.
\eqno(A.8)
$$
Noting that $u(p,\lambda)$ and $v(p,\lambda)$ are the solutions of Dirac 
equations, one finds that $C$ must fulfill the following conditions,
$$
   C\gamma_\mu^T C^{-1} = -\gamma_\mu.
\eqno(A.9)
$$
This condition is obtained also from the requirement that
$$
    (i\slash \partial - e\slash A - m) \Psi(x) =0,
\eqno(A.10)
$$
leads to the following equation of motion for $\Psi^c$,
$$ 
    (i\slash \partial + e\slash A - m) \Psi(x)^c =0.
\eqno(A.11)
$$
 From the consistency between two equations in (A.5) or in (A.6), and the
unitarity of $C$, $C^\dagger = C^{-1}$, one finds again (A.8).\par
    The Majorana particle is characterized by the condition ,
$$
    \Psi=\Psi^c.
\eqno(A.12)
$$
Consequently, the Majorana field $\Psi$ is  expanded as 
$$
\eqalign{
    \Psi(x) =& \sum_\lambda \int{{d^3p}\over{(2\pi)^3 2p_0}}
    [c(p\lambda)u(p\lambda)e^{-ipx}+c^\dagger(p\lambda)v(p\lambda)e^{ipx}],\cr
 \bar\Psi(x) =& \sum_\lambda \int{{d^3p}\over{(2\pi)^3 2p_0}}
    [c(p\lambda)\bar v(p\lambda)e^{-ipx}+c^\dagger(p\lambda)\bar u(p\lambda)
        e^{ipx}],\cr
}\eqno(A.13)
$$

There is no vector current nor tensor current for Majorana particles:
$$
   \bar\Psi_M\gamma_\mu\Psi_M = \bar\Psi_M\sigma_{\mu\nu}\Psi_M =0,
\eqno(A.14)
$$
since 
$$
    C\gamma^T_\mu C^{-1} = -\gamma_\mu,~~~~~
    C\sigma^T_{\mu\nu} C^{-1} = -\sigma_{\mu\nu}.
\eqno(A.15)
$$
\vfill\eject

\noindent{\bf Appendix B. Summary of Grace line rule}\par
\item{1.} Fix the underlying theory and determine the model 
lagrangian which defines the vertex functions,
$$
       \Gamma_{\xi,\eta}=\xi \leftarrow \bullet\leftarrow \eta
$$
with $(\xi,\eta)= (D_i,D_j), (M_\alpha,D_i), (D_i,M_\alpha), (C_i,D),
(D,C_i)$ and $(M_\alpha,M_\beta)$ with $\alpha>\beta$. 
Only these vertex functions appear in the Grace line amplitudes.  In 
particular, the lagrangian for Majorana-Majorana vertex must be brought 
in the form of (27).
\item{2.} Assign particle numbers 1 to $n_i$ to
$n_i$ incoming particles (in the order of Dirac, anti-Dirac and Majorana
particles) and $n_i+1$ to $n_i+n_f$ to $n_f$ outgoing particles 
(in the order of anti-Dirac, Dirac and Majorana particles).
\item{3.}   Draw Feynman diagrams for a given process and the order of
perturbation based on the underlying model lagrangian.
\medskip
For each Feynman diagram:\par
\item{4.} Read out  $P=(a_1,a_2)(a_3,a_4).......(a_{ni+nf-1},a_{ni+nf})$ from
the diagram and determine $\Sign(P)$. Note that the pair $(m,n)$ must be 
arranged such that  $m>n$.  If there are several Feynman diagrams which give
the same $P$,  count and treat them separately.
\item{5.} Draw ${{n_i+n_f}\over 2}$ Grace lines $G_{m,n}$ corresponding
to the pair $(m,n)$ with $m>n$.\par
\medskip
For each Grace line $G_{m,n}$:\par
\item{6.}Assign the F-direction and M-direction to each segment of the Grace 
line based on the rule given in section 3.  The M-direction is the same as the
physical momentum direction for external particles, and for internal particles
it is in the same direction as the A-direction. The F-direction of the Dirac
particle is defined by the direction of fermion number flow.  The F-direction
of Majorana particles is in the same direction of their M-direction.
\item{7.} For external spinors, assign spinor wave functions  based on the
rule given in  Table 2.
\item{8.}Put ($-$) for each external spinor when its M-direction is opposite
to the A-direction.
\item{9.}  Use the propagator $S(k)$ or $-S(-k)^T$, depending on whether
the F-direction is in or opposite to the A-direction.
\item{10.}  Use the vertex function given by (18) and (19) for
Dirac-Dirac vertices, (28) and  (29) 
for Dirac-Majorana vertices, (30)-(43) for Majorana-Majorana vertices
and (46) for the vertices including charginos.
\medskip
\item{11.}  Finally, multiply ${{n_i+n_f}\over 2}$ Grace line 
amplitudes $[G_{m,n}]$ and the appropriate bosonic parts which 
connect vertices on the Grace lines in order to obtain 
an amplitude for each Feynman diagram, and sum  up the resulting
amplitudes over all the possible  partitions $P$ (corresponding
to all Feynman diagrams):  
$$ {\cal M} = \sum_P \Sign(P){\cal M}_P,~~~~~
   {\cal M}_P =\Pi_{(m,n)}[G_{m,n}]({\rm bosonic~part})_{m,n}.
$$
     
\vfill\eject

\noindent{\bf Appendix C. Grace line rule without charge conjugation matrix}
\par
\item{1.} Fix the underlying theory and determine the model 
lagrangian which defines the vertex functions,
$$
       \Gamma_{\xi,\eta}=\xi \leftarrow \bullet\leftarrow \eta
$$
with $(\xi,\eta)= (D_i,D_j), (M_\alpha,D_i), (D_i,M_\alpha), (C_i,D), 
(D,C_i)$ and $(M_\alpha,M_\beta)$ with  $\alpha > \beta$. 
For Majorana-Majorana vertex, bring the lagrangian in the form of (27) and
neglect the F-direction of Majorana lines.
\item{2.} Prepare $\Gamma^c\equiv C\Gamma^TC^{-1}$ for each vertex function
$\Gamma$.  Note that\par
~~~~$
\Gamma^c=(1,~\gamma_5,~-\gamma_\mu,~ \gamma_\mu\gamma_5,~-\sigma_{\mu\nu})$
~~~~for~~~~ 
$\Gamma=(1,~\gamma_5,~\gamma_\mu,~ \gamma_\mu\gamma_5,~\sigma_{\mu\nu}).
$
\item{3.} Assign particle numbers 1 to $n_i$ to
$n_i$ incoming particles (in the order of Dirac, anti-Dirac and Majorana
particles) and $n_i+1$ to $n_i+n_f$ to $n_f$ outgoing particles 
(in the order of anti-Dirac, Dirac and Majorana particles).
\item{4.}   Draw Feynman diagrams for a given process and the order of
perturbation based on the underlying model lagrangian.
\medskip
For each Feynman diagram:\par
\item{5.} Read out  $P=(a_1,a_2)(a_3,a_4).......(a_{ni+nf-1},a_{ni+nf})$ from
the diagram and determine $\Sign(P)$. Note that the pair $(m,n)$ must be 
arranged such that  $m>n$.  If there are several Feynman diagrams which give
the same $P$,  count and treat them separately.
\item{6.} Draw ${{n_i+n_f}\over 2}$ Grace lines $G_{m,n}$ corresponding
to the pair $(m,n)$ with $m>n$.\par
\medskip
For each Grace line $G_{m,n}$:\par
\item{7.}Assign the M-direction to each segment of the Grace 
line based on the rule given in section 3.  The M-direction is the same as the
physical momentum direction for external particles, and 
it is in the same direction as the A-direction for internal particles. 
\item{8.} Assign the F-direction to each segment of the Grace line.  First,
assign the F-direction to Dirac particles, which is taken in the direction 
of fermion number flow. Then assign the F-direction to Majorana particles 
starting 
from the right endpoint, so that the F-directions do not clash as much as 
possible( that is, until the Dirac particle propagator which has a different
F-direction from that of right-end spinor appears.) Repeat the procedure
for the next Majorana propagator.  If the right endpoint spinor is
Majorana particle, start with the left endpoint Dirac spinor.  If
both of the endpoint spinors are Majorana particles, start with a Dirac
spinor in the propagator.  If there is no Dirac spinor in a Grace line,
assign
the F-direction of the Grace line along the A-direction. 
\item{9.} For external spinors, assign spinor wave functions  based on the
rule given in  Table 4.
\medskip
 
{\offinterlineskip \tabskip=0pt
\halign{ \strut
  \vrule#& \quad # \quad &
  \vrule#& \quad # \quad &
  \vrule#& \quad # \quad &
  \vrule#& \quad # \quad &
  \vrule#& \quad # \quad &
  \vrule#  \cr
\noalign{\hrule}
  & F-direction  && $\longleftarrow\bullet$ &&  $\longrightarrow\bullet$ &&
                    $\bullet\longleftarrow$ &&  $\bullet\longrightarrow$ & \cr
  &  M-direction &&                 &&               && &&              & \cr
\noalign{\hrule}
  & $\longleftarrow$  &&   $~\bar u(p)$   &&    $ \bar u(p)$  
                      &&   $u(p)$         &&    $ u(p)$ & \cr
  & $\longrightarrow$ &&   $~\bar v(p)$   &&    $ \bar v(p)$  
                      &&   $v(p)$         &&    $ v(p)$ & \cr
\noalign{\hrule} }} 
\medskip
\noindent{ Table 4:} Spinor assignment at left endpoint and at right endpoint.
\medskip

\item{10.} For propagators, use $S(k)$ independent of the F-directions of the
spinors.  For the vertex function, use $\Gamma_{\xi,\eta}$ for 
$\xi\leftarrow\bullet\leftarrow\eta$ and $\Gamma^c_{\eta,\xi}= C
\Gamma^T_{\xi,\eta}C^{-1}$ for $\eta\rightarrow\bullet\rightarrow\xi$.  
For clashing vertices, use $\hat \Gamma$,
where the rules (60) and (61) are to be applied. By construction 
there are no clashing vertices for $(M_\alpha, M_\beta)$.
\item{11.} Multiply the Grace line amplitude so constructed by factor 
$(-)^{V_g}$ where $V_g$ is the number of $v$ and $\bar v$ appearing in the 
Grace line.  This procedure can be neglected since $\sum V_g$ is common
to all Feynman diagrams in a given process.  Here, the sum is taken 
over the Grace lines in the
Feynman diagram we are concerned with.
\medskip
\item{12.}  Finally, multiply ${{n_i+n_f}\over 2}$ Grace line 
amplitudes $[G_{m,n}]$ and the appropriate bosonic parts which 
connect vertices on the Grace lines in order to obtain 
an amplitude for each Feynman diagram, and sum  up the resulting
amplitudes over all the possible  partitions $P$ (corresponding
to all Feynman diagrams):  
$$ {\cal M} = \sum_P \Sign(P){\cal M}_P,~~~~~
   {\cal M}_P =\Pi_{(m,n)}[G_{m,n}]({\rm bosonic~part})_{m,n}.
$$

 \end
 
 \end